# Demonstration of weak measurements, projective measurements, and quantum-to-classical transitions in ultrafast free electron-photon interactions

Yiming Pan[1,6†], Eliahu Cohen[2†], Ebrahim Karimi[3], Avraham Gover[4], Norbert Schönenberger[5], Tomáš Chlouba[5], Kangpeng Wang[6], Saar Nehemia[6], Peter Hommelhoff[5], Ido Kaminer[6] and Yakir Aharonov[7,8]

1. Department of Physics of Complex Systems, Weizmann Institute of Science, Rehovot 7610001, Israel
2. Faculty of Engineering and the Institute of Nanotechnology and Advanced Materials, Bar Ilan University, Ramat Gan 5290002, Israel
3. Department of Physics, University of Ottawa, Ottawa, Ontario, K1N 6N5, Canada
4. Department of Electrical Engineering Physical Electronics, Tel Aviv University, Ramat Aviv 6997801, Israel
5. Department of Physics, Friedrich-Alexander Universität Erlangen-Nürnberg (FAU), Staudtstraße 1, 91058 Erlangen, Germany
6. Department of Electrical Engineering, Technion, Haifa 3200003, Israel
7. School of Physics and Astronomy, Tel Aviv University, Ramat Aviv 6997801, Israel
8. Institute for Quantum Studies, Chapman University, Orange, CA 92866, USA

†Correspondence and requests for materials should be addressed to E.C. (eliahu.cohen@biu.ac.il) and Y.P. (yiming.pan@campus.technion.ac.il).

**Abstract**

How does the quantum-to-classical transition of measurement occur? This question is vital for both foundations and applications of quantum mechanics. Here, we develop a new measurement-based framework for characterizing the classical and quantum free electron-photon interactions and then experimentally test it. We first analyze the transition from projective to weak measurement in generic light-matter interactions and show that any classical electron-laser-beam interaction can be represented as an outcome of a weak measurement. In particular, the appearance of classical point-particle acceleration is an example of an amplified weak value resulting from weak measurement. A universal factor, $\exp(-\Gamma^2/2)$, quantifies the measurement regimes and their transition from quantum to classical, where $\Gamma$ corresponds to the ratio between the electron wavepacket size and the optical wavelength. This measurement-based formulation is experimentally verified in both limits of photon-induced near-field electron microscopy and the classical acceleration regime using a dielectric laser accelerator. Our results shed new light on the transition from quantum to classical electrodynamics, enabling to employ the essence of wave-particle duality of both light and electrons in quantum measurement for exploring and applying many quantum and classical light-matter interactions.

## Introduction

Measurement lies at the heart of quantum mechanics and allows one to probe a quantum system of interest through a measuring pointer (an apparatus) coupled to the system's observables. The interaction between the system and pointer is later classically amplified for the outcome to be seen macroscopically. However, in the context of light-matter interactions, sometimes either the measured system or measuring pointer (or both) can be well treated with classical means, i.e., without invoking the quantum formalism. These interactions are usually modeled by classical or quantum electrodynamics, with a wealth of widely explored effects and both theoretical and experimental schemes such as photon-induced near-field electron microscopy (PINEM) [1,2] or dielectric laser accelerator (DLA) [3,4]. All these inspire our current exploration. We show that there is a continuous transition from quantum to classical interactions between electrons and photons, which can be illustrated when examining several limiting cases of the measurement process in actual experimental setups.

We wish to investigate the various regimes when electrons and photons are coupled, to classify in which cases they can be regarded as 'classical' or 'quantum' measuring pointers. In particular, we study the transition process between the two regimes. In light of the current experimental capabilities of manipulating electrons and photons, the quantitative wave-particle duality of electrons and also the quantum-to-classical transitions of photons are both controllable in ultra-fast transmission electron microscopy (UTEM) [1,2] and in quantum light preparation [5], respectively. In the wavepacket representation with electron wavepacket size ($\Delta_z$), the point-particle-like ('classical') picture of free electrons can be defined in the limit $\Delta_z \to 0$ and conversely, the plane-wave-like ('quantum') picture in the opposite limit $\Delta_z \to \infty$. Similarly, the photon state holds its own quantum-to-classical transition. For concreteness, the single-photon-added coherent state enables us to continuously tune the photon system from a coherent state (representing quantum states in their 'classical' limit) to a single Fock number state (which we take as a uniquely 'quantum' state) [5,6]. Note that indeed, the coherent state delineates the border regarding the classicality and quantumness of photon states from different perspectives [7]. We thus define a parameterized photon state as the basis for possible investigation of the fuzzy border that may separate the 'quantum' from 'classical' regimes in the above sense, utilizing the coupling with a single electron wavepacket as a measuring pointer *[Footnote 1: We note that the above* *characterization of classical and quantum states is explicitly tailored to the analysis of interactions between free electrons and photons. Indeed, other notions of classicality could be found in literatures.]*

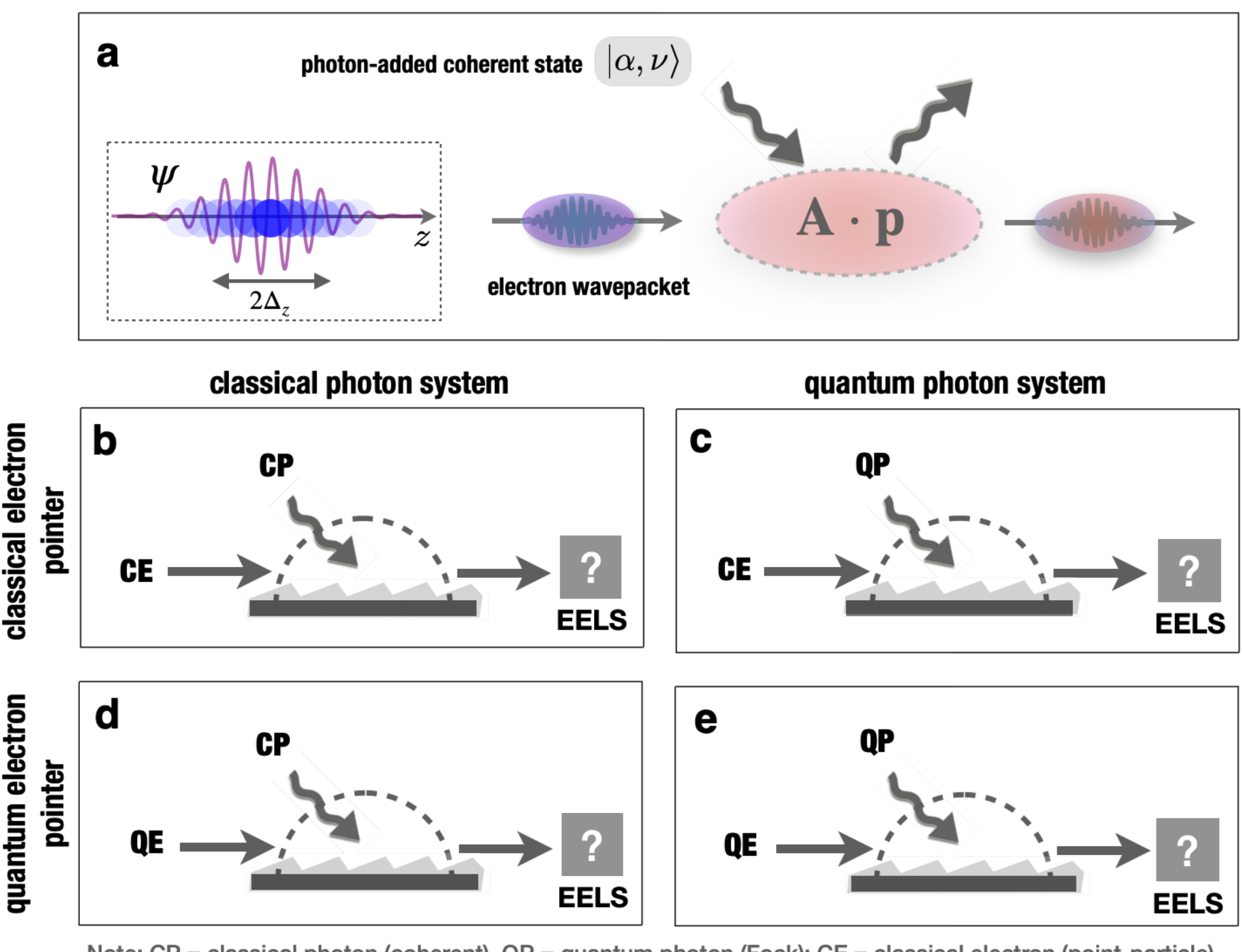

**Fig.1: The quantum and classical measurement schemes of electron-photon interactions.** (a) The classical photon (CP) in a coherent state and quantum photon (QP) in a Fock state are defined as two opposite limits of the photon-added coherent state $|\alpha,\nu\rangle$, where $\nu = 0$ and $\alpha = 0$ respectively. The classical electron (CE) and quantum electron (QE) are defined as the wavepacket representation in the point-particle limit and plane-wave limit, respectively (Eq. (1) in the text). The measuring pointer is the outgoing electron, and the system is the pre-prepared photon state (without post-selection), with a coupling strength g between the system and pointer. (b-e) Four combinations of electron and photon interactions are presented in the classical and quantum

measurement regimes. The readout of the measuring pointer is the electron energy loss spectrum (EELS).

To be specific, we represent the 'quantum-to-classical' transition of photon states using the photon-added coherent state and the 'particle-to-wave' electron state using a Gaussian wavepacket. The initially prepared photon and electron states are respectively given by

$$|\alpha,\nu\rangle = \frac{(a^{\dagger})^{\nu}|\alpha\rangle}{\sqrt{\nu!\,L_{\nu}(-|\alpha|^{2})}},$$
$$|\psi\rangle = \int dp\, c_{p}^{(0)}\,|p\rangle. \tag{1}$$

The photon-added coherent state reduces to the limit of Fock or coherent state for the parameters $\alpha \to 0$ or $\nu \to 0$, respectively, with $L_{\nu}$ being the Laguerre polynomial of (integer) order $\nu$, $a^{\dagger}$ is the photon creation operation, and all other photon indices are suppressed for simplicity. Such a photon state was theoretically proposed by Agarwal and Tara [6] and later experimentally realized by Zavatta *et al.* [5]. The normalized Gaussian component of a free electron wavefunction is

$$c_{p}^{(0)} = \left(2\pi\Delta_{p}^{2}\right)^{-1/4}\exp(-(p-p_{0})^{2}/4\Delta_{p}^{2})$$

with $\Delta_{\mathrm{p}}$ the unchirped momentum uncertainty, and $p_{0}$ the effective electron mass, the velocity, the average momentum, and energy, respectively. Note that the electron wavepacket is only defined in a longitudinal dimension (1D), where the electron's initial momentum distribution is readily obtained as $\rho^{(0)}(p) = \left|c_{p}^{(0)}\right|^{2}$. Following the standard procedure of measurement proposed by von Neumann [8], we can study the quantum-to-classical transitions of measurement of free electron-photon coupling as a testing platform of the system-pointer measurement schemes by classifying it into four types of interaction, as shown in Figs. 1b-e: (I) the classical point-particle electron coupling with 'classical' photon coherent state; (II) the classical point-particle electron coupling with 'quantum' photon Fock state; (III) the quantum plane-wave electron coupling with classical photon; (IV) the quantum plane-wave electron coupling with quantum photon.

Next, we assume that the coupling between the classical electron ($\Delta_z = \frac{\hbar}{\Delta_p} \to 0$) and the classical photon ($\nu \to 0$ but $\alpha \neq 0$) can be simplified into the canonical Hamilton equations $\dot{z} = p/\gamma_0 m, \dot{p} = eE_c \cos(\omega t - q_z z(t) + \phi_0)$, which in Newtonian mechanics describe a charged point-particle ($-e$) moving in the presence of a monochromatic traveling electromagnetic field (laser, or microwave field) with electric component $E = E_c \cos(\omega t - q_z z(t) + \phi_0)$ having the optical frequency ω and the z component of the wave vector $q_z$ along the propagation direction. With the short-time approximation $z(t) = v_0 t$, we expect that the point-particle momentum transfer can be thus reduced to $\Delta p_{point} = -eE_c L/v_0 \operatorname{sinc}\left(\frac{\bar{\theta}}{2}\right) \cos\left(\frac{\bar{\theta}}{2} + \phi_0\right)$, in which the synchronization condition (also, called phase matching condition) is $\bar{\theta} = (\omega/v_0 - q_z)L$, $L$ is the interaction length and $v_0$ is the initial velocity of the electron. This is the well-known linear acceleration formula in classical accelerator physics, as well as in the inverse Smith-Purcell effect, or the Dielectric Laser Accelerator (DLA) [3,4] and free electron lasers [9]. In addition, it indicates that the emergence of 'classicality' in our measurement setup requires both the classical conditions of 'point-particle-like' electron and photon at a coherent state, as shown in Fig. 1b.

This classical acceleration formula offers a hint how to quantum-mechanically measure the electromagnetic field operators (e.g., the vector potential **A**) via a moving electron wavepacket as a measuring pointer coupled to the measured photonic system. It will be shown how to calculate the classical particle acceleration within the von Neumann measurement scheme [8], as a result of the electron-photon coupling. From the perspective of weak measurement [10], we will see below that the momentum transfer of the pointer after interaction corresponds to the weak value of the vector potential (**A**) of the photonic system. This applies to the configuration of classical electron pointer coupled to a classical photon system (Fig. 1b). In the other three configurations, the electron-photon couplings indicate a quantum (strong) projective measurement. As a result, the system-pointer measurement inevitably falls into the 'strong' category involving a significant momentum change with a subsequent 'wavefunction collapse', regardless of whether the electron or photon state falls in the quantum regime (Fig. 1c-e). The classification of four measurement regimes will indicate in the following sections how only quantum weak measurement can lead to the classical acceleration (Figs. 1b and 2a), thereby possibly implying in general how classical electrodynamics may emerge from a full quantum treatment.

Additionally, in the transition from weak to projective measurements, we notice that the identities of electron and photon are reciprocal in the following sense: which is the system and which is the pointer depends on the detection and post-selection configuration of electrons and photons. This underlying reciprocity leads to the system-pointer duality that will be discussed towards the end of this work.

The main novelty of the work relies in the presentation of a unified scheme for analyzing interactions between photons and free-electrons addressing both their classical and quantum regimes, as well as the interesting quantum-to-classical transition. The proposed scheme employs tools from quantum measurement theory and therefore makes it easy to draw the line between weak and strong (projective) measurements in light-matter interactions. In particular, two limits of our theoretical prediction are experimentally tested here with recent UTEM and DLA setups.

The rest of the manuscript is organized as follows. In the next two sections we differentiate between the four basic types of electron-photon couplings, addressing inter alia the emergence of point-particle trajectories. We then analyze the expectation-valued electron spectrum and weak-valued spectrum in our measurement transition theory and find that the classical electron momentum transfer can be viewed as an outcome of a weak measurement in the next two sections. We then verify our theoretical predictions using experimental results obtained in UTEM and DLA setups. Finally, we address the advances of post-selection in light-electron interactions, and further clarify the significance of classical-to-quantum transitions in our measurement-based theory.

**Classical photon in a coherent state**. Our analysis of measurement is based on the perturbative solution of the relativistically modified Schrödinger equation [9, 11, 12] for a free electron wavefunction and a quantized radiation field. Following the standard QED treatment (see the SI file) we expand the initial wavefunction in terms of the quantum continuous numbers $p$ of the electron state and the Fock number-occupation state of the photon, which is given by $|i\rangle = \sum_{p,\nu} c_{p,\nu}^{(0)} e^{-iE_p t/\hbar}|p,\nu\rangle$, where $c_{p,\nu}^{(0)}$ is the component of the combined basis $|p,\nu\rangle = |p\rangle \otimes |\nu\rangle$.

To access energy transfer in electron spectrum, the wavepacket acceleration as the pointer shift is obtained in first order perturbation theory as $\Delta E = \sum_{p,\nu} \left| c_{p,\nu}^{(0)} + c_{p,\nu}^{(1)(e)} + c_{p,\nu}^{(1)(a)} \right|^2 \left(E_p - E_0\right)$ where the initial electron energy $E_0 = \sum_{p,\nu} \left| c_{p,\nu}^{(0)} \right|^2 E_p$ [10-12]. In our quantum treatment of the initial electron-photon state as given by $c_{p,\nu}^{(0)} = c_p^{(0)} c_\nu^{(0)}$, we consider the initial electron wavepacket of the unchirped Gaussian distribution (Eq. 1) combined with a coherent photon state, where $\nu_0 = \sum_\nu \nu \left| c_\nu^{(0)} \right|^2$ is the total photon number. Substituting into the acceleration formula ($\Delta E$), one can obtain the explicit energy transfer with two parts ($\Delta \mathrm{E} = \Delta \mathrm{E}^{(1)} + \Delta \mathrm{E}^{(2)}$) [9]:

$$
\begin{aligned}
\Delta E^{(1)} &= \Delta \mathrm{E}_{\mathrm{point}} e^{-\Gamma^2/2} \\
\Delta E^{(2)} &= -\widetilde{\Upsilon}^2 \hbar\omega \, sinc^2 \left( \frac{\bar{\theta}}{2} \right)
\end{aligned}
\tag{2}
$$

where $\Delta \mathrm{E}_{\mathrm{point}} = \nu_0 \Delta \mathrm{p}_{\mathrm{point}} = -\mathrm{eE_c L} \, sinc \left( \frac{\bar{\theta}}{2} \right) \cos \left( \frac{\bar{\theta}}{2} + \phi_0 \right)$ and the normalized photon exchange coefficient of spontaneous emission is defined as $\widetilde{\Upsilon} = e\tilde{E}_q L / 4\hbar\omega$, corresponding to the PINEM near-field factor as $g = 2\sqrt{\nu_0}\, \widetilde{\Upsilon}$ (see the SI file). Importantly, the relation $\sqrt{\nu_0} = \langle \sqrt{\nu_0} | a_\nu | \sqrt{\nu_0} \rangle$ is taken for the coherent state ($|\sqrt{\nu_0}\rangle$). A significant pointer-specific extinction parameter $e^{-\Gamma^2/2}$ is found in the phase-dependent energy transfer (2), with a decay parameter given by

$$
\Gamma = \frac{2\pi}{\beta} \left( \frac{\Delta_z}{\lambda} \right) = \frac{\hbar\omega}{\nu_0 \Delta_{\mathrm{p}}}
\tag{3}
$$

with $\beta = v_0/c$. The extinction parameter demonstrates that it is the pre-interaction history-dependent wavepacket size of a free-electron wavepacket that has physical effect in its interaction with coherent light and acts as the measuring pointer.

Now, we are able to discuss quantitatively the classical point-particle and quantum plane-wave limits of electron wavepacket acceleration, in the interaction with quantized photon state of light, as shown in Figs. 1b,d. The particle-to-wave transition of the electron-photon interaction in measuring electron energy loss spectroscopy (EELS) is shown in Fig. 2a. The appearance of 'classicality' corresponds to the case where the photon distribution becomes a coherent state describing the 'classical' electromagnetic field and the condition $e^{-\Gamma^2/2} \to 1$ is satisfied, which means that the electron wavefunction looks like a point-like particle with wavepacket size smaller than the wavelength: $\Delta_z \ll \lambda$ ("narrow" electron). Indeed, the wavepacket-dependent acceleration when the wavepacket size is comparable to the wavelength is given by $\Delta E = \Delta \mathrm{E}_{\mathrm{point}} e^{-\Gamma^2/2}$. Therefore, it explains the emergence of classical point-particle trajectory in the free electron-photon setup of "CE + CP". The decay parameter ($\Gamma$) implies the measurability of the electron wavepacket size near the classical particle-like regime.

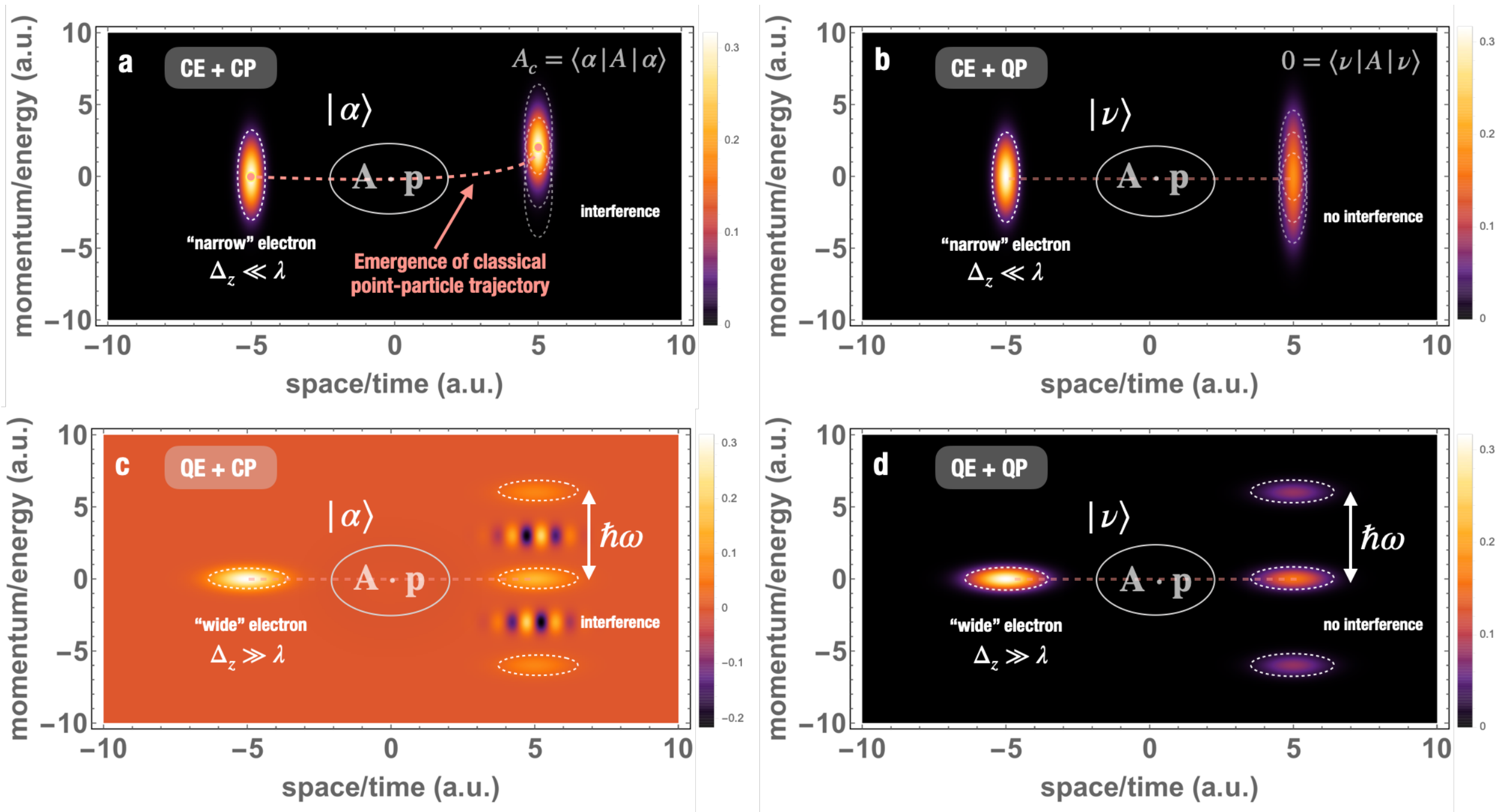

**Fig. 2: Illustration of four measurement regimes of free electron-photon interactions in phase-space representation.** The electrons are presented using Wigner functions corresponding to the specified measurement cases as shown in Fig. 1. (a) classical electron (CE) interacts with a classical photon (CP) ($\Delta_z \ll \lambda$, coherent state $|\alpha\rangle$); (b) CE interacts with a quantum photon (QP) ($\Delta_z \ll \lambda$, Fock state $|\nu\rangle$); (c) quantum electron (QE) interacts with a CP ($\Delta_z \gg \lambda, |\alpha\rangle$); (d) QE interacts with a QP ($\Delta_z \gg \lambda, |\nu\rangle$). Crucially, among them only the case (a) "CE + CP" gives rise to the weak measurement and classical point-particle trajectory in electrodynamics. One can obtain the net wavepacket acceleration as a weak value and observe the emergent classical electron dynamics obeying a classical trajectory with certain position and momentum in phase space, as shown in (a).

On the other hand, the plane-wave limit can be directly defined as $e^{-\Gamma^2/2} \to 0$ in which case $\Delta E^{(1)} \to 0$, and the energy transfer has only the contribution of phase-independent term ($\Delta E^{(2)}$), as shown in Fig. 2c. Note that even in the classical limit, the phase-independent term still has a universal non-vanishing noise contribution in the form of vacuum fluctuations. This phase-independent term ($\Delta E^{(2)}$) relates to the vacuum expectation value, which acts as quantum noise of spontaneous fluctuation in our electron-photon coupling measurements [9]. Therefore, the phase-dependent term ($\Delta E^{(1)}$) reduces to the classical particle acceleration but is measurable only if the spontaneous vacuum fluctuation is negligible: $\Delta E^{(1)} \gg \Delta E^{(2)}$ under $\nu_0 \gg 1$.

**Quantum photon in a Fock state**. In contrast, the single Fock state of light corresponds to the photon-added coherent state (Eq. 1), obeying the condition $\alpha \to 0$, i.e., $c_{\nu}^{(0)} = \delta_{\nu,\nu_0}$. When inspecting the wavepacket acceleration expression, it appears that similarly to the case of spontaneous emission, there is no Fock state stimulated energy transfer due to the orthogonality relations $\langle \nu_0 | a_\nu | \nu_0 \rangle = \langle \nu_0 | a_\nu^\dagger | \nu_0 \rangle = 0$. Therefore, one obtains the total energy transfer, $\Delta \mathrm{E} = \Delta \mathrm{E}^{(1)} + \Delta \mathrm{E}^{(2)} = -\widetilde{\Upsilon}^2 \hbar\omega \, sinc^2 \left(\frac{\overline{\theta}}{2}\right)$. There is no stimulated radiative interaction as a result of the coupling to the quantum light (radiation wave) in Figs. 1c,e. However, this is not very surprising since the initial single Fock state ($|\nu_0\rangle$) is orthogonal to the emitted and absorbed photon state ($|\nu_0 \pm 1\rangle$), so that the first order phase-dependent interference term has no contribution. Therefore, the phase-space descriptions of an electron interacting with a quantum light source, depicted in Figs. 2b,d present either momentum broadening (in the classical electron case) or distinct quantum sidebands (in the quantum electron case), without the contribution of quantum interference between photon sidebands depicted in Fig. 2c in the case of classical light quantum electron case [13]. Note that the second term ($\Delta \mathrm{E}^{(2)}$) still produces the wavepacket-independent spontaneous vacuum fluctuations as an inevitable source of quantum noise in the observation of EELS in the quantum light case, same as in the case of classical light (Eq. 2).

**Weak measurement versus projective measurement**. Let us focus now on the EELS observation of the final electron wavefunction after the interaction. When a quantum electron pointer is coupled to the photon system, the photon-induced outgoing electron momentum distribution is then given by integrating out all photonic degrees of freedom, that is, $\rho^{(f)}(p) = \sum_{\nu} \left| c_{p,\nu}^{(0)} + c_{p,\nu}^{(1)(e)} + c_{p,\nu}^{(1)(a)} \right|^2$. Let us find the EELS measurement pictures in the two aforementioned limits.

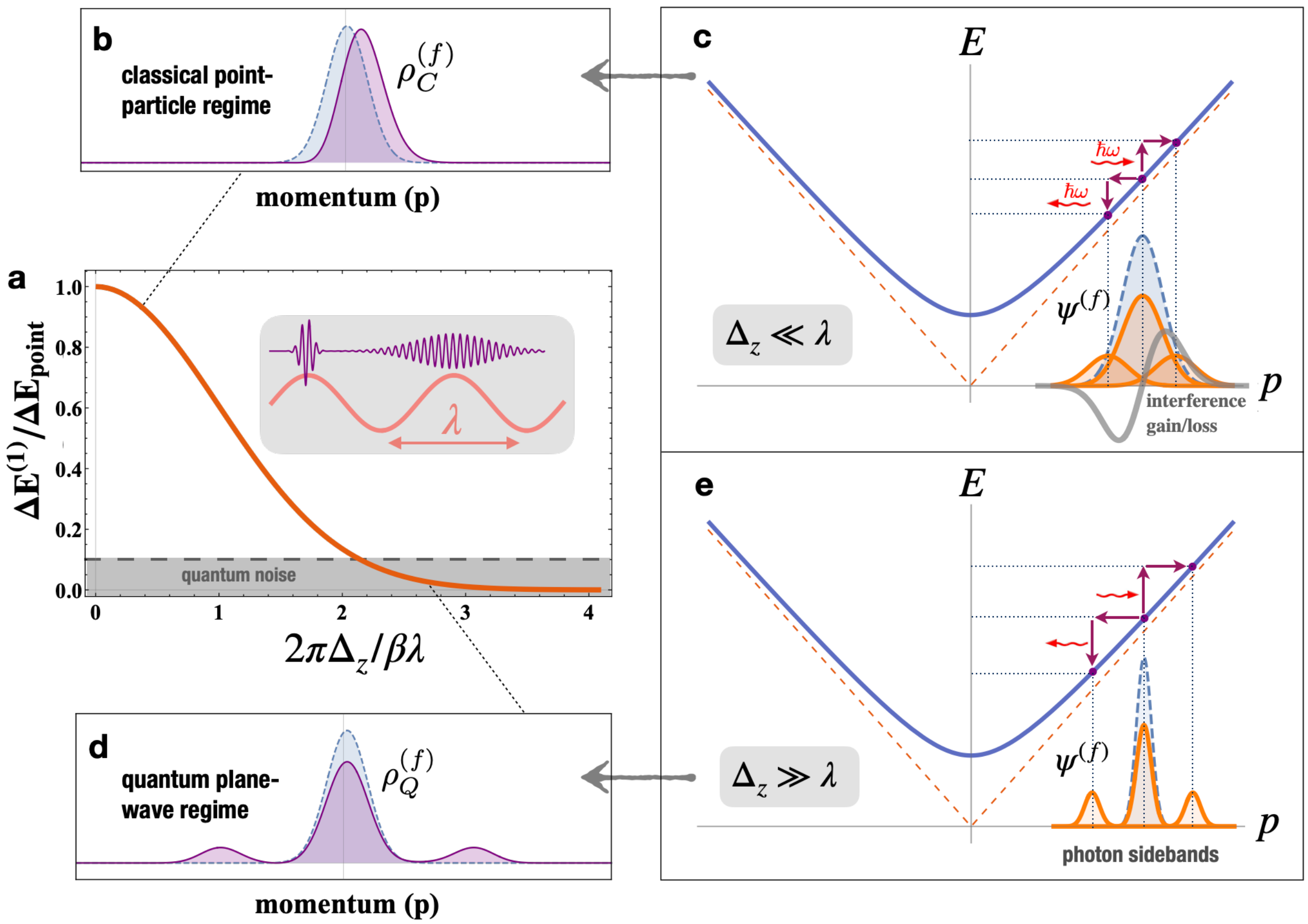

**Fig. 3: Quantum-to-classical measurement transition** between the classical point-particle picture (b, c) and quantum plane-wave picture (d, e) of electron wavepacket pointer when coupled to a photon coherent state. The two limits of particle-like (b) and wave-like (d) pictures of the electron pointer in light-matter interaction corresponds to the classical (weak) measurement (particle acceleration) and quantum (projective) measurement (PINEM), respectively. The exact expressions of the final electron momentum distributions are presented in the text.

First, in the point-particle limit $\Delta_z \ll \lambda$ [9, 14], necessarily the initial momentum distribution exceeds the quantum momentum recoil $\Delta_p > \hbar\omega/v_0$ and hence the final momentum distribution after interaction with the classical photon is: $\rho_C^{(f)}(p) = \rho^{(0)}(p - \Delta p^{(1)})$, where the momentum shift is $\Delta p^{(1)} = \Delta \mathrm{p}_{\mathrm{point}} e^{-\Gamma^2/2}$ (also corresponding to the energy transfer $\Delta \mathrm{E}^{(1)}$in Eq. 2). As shown in Figs. 3b,c, the emission and absorption terms overlap with the initial wavepacket momentum distribution and contribute the asymmetrical interference effects with opposite sign, which leads to the momentum shift in the classical point-particle regime. The final momentum distribution of the electron pointer is then reshaped, displaying a net momentum shift of small acceleration, as shown in Figs. 3b,c, where we ignore the spontaneous term in the weak-field coupling $eE_cL/\hbar\omega < 1$. Except for the universal transition factor $e^{-\Gamma^2/2}$, the acceleration/deceleration of the electron wavepacket depends on the synchronism detuning parameter $\overline{\theta}$ and the relative phase $\phi_0$ , similar to a charged point-particle moving in the presence of a classical electromagnetic field ($\Delta \mathrm{p}_{\mathrm{point}}$) in the classical limit of interaction between 'particle-like' electron and 'classical' photon (Figs. 1b and 2a). This interaction picture of electron-photon coupling leads to the classical measurement or classical electrodynamics, and also to the weak measurement as displayed in Fig. 3b.

Next, in the plane-wave (quantum electron) limit $\Delta_{\mathrm{z}} \gg \lambda$, corresponding to the large recoil condition $\Delta_{\mathrm{p}} < \hbar\omega/v_0$ (i.e., the criterion of projective measurement), the interference terms between sidebands diminish, and the scattered components dominate, resulting in a final PINEM-kind spectrum of the momentum distribution, as shown in Figs. 3d,e: $\rho_Q^{(f)}(p) = \left(1 - 2\Upsilon^2 sinc^2\left(\frac{\overline{\theta}}{2}\right)\right)\rho^{(0)}(p) + \Upsilon^2 sinc^2\left(\frac{\overline{\theta}}{2}\right)\left(\rho^{(0)}(p - \hbar\omega/v_0) + \rho^{(0)}(p + \hbar\omega/v_0)\right)$ where $\Upsilon = eE_cL/4\hbar\omega$ and we ignore the spontaneous contribution to the emission term with the approximation $v_0 \gg 1$. The last two scattering terms represent symmetric photon-sideband spaced by $\hbar\omega/v_0$on both sides of the central momentum $p_0$ of the wavepacket as displayed in Fig. 3e. This quantum measurement result is the same as the multiple sidebands electron energy gain/loss spectrum in PINEM experiments, in which the higher-order sidebands relate to multiple-photon emission and absorption processes [1-2].

For the Fock state of the photon system, the phase-dependent interference terms disappear due to the orthogonality, and thus lead to the same final projective momentum distribution as the measurement in the plane-wave (classical light) limit, regardless of the electron's wavefunction profile in the classical or quantum limit. For the other three electron-photon couplings in Fig. 1c-e, either quantum electron or quantum photon corresponds to the final projective momentum distribution with no net momentum transfer $\Delta p = \int \rho^{(f)}(p) p \, dp = 0$ (i.e., $\Delta \mathrm{E} = 0$), which implies no classical measurement for these three system-pointer interaction configurations.

**Is the wavepacket acceleration a weak value?** As demonstrated in Figs. 3b,d, we find that the projective measurement [8] corresponds to the electron spectrum with discrete photon-sidebands of PINEM (3d), and the weak measurement [10] to the acceleration spectrum with central momentum shift (3b). Moreover, the energy/momentum transfer is proportional to the classical electric field given by $\boldsymbol{E} = -\langle \partial \boldsymbol{A} / \partial t \rangle$ in the Coulomb gauge $\nabla \cdot \boldsymbol{A} = 0$. Our results seem depending on the gauge choice of vector potential $\boldsymbol{A}$, but in our case, it is just $\boldsymbol{A} = \int \boldsymbol{E} \, dt$, i.e., completely defined by the physically gauge-independent electric field (up to a meaningless integration constant). Such gauge-independence was shown to arise when performing a weak measurement of the vector potential and measuring the Berry phase [15, 16]. Thus, the classical point-particle acceleration is an effective weak value of the vector potential in the formalism of weak measurement,

$$\Delta \mathrm{p}_{\text{point}} \propto A_w \equiv \frac{\langle \beta, \nu' | \boldsymbol{A} | \alpha, \nu \rangle}{\langle \beta, \nu' | \alpha, \nu \rangle} \tag{4}$$

where the pre- and post-selected photon states are both defined as photon-added coherent states (1). Note that this definition of the vector-potential weak value is applicable only if there is no time evolution of the photon system (except for the measurement process), or effectively, in short-time approximation. Two typical examples are considered with fixing the pre-selection and post-selection at the 'classical' or 'quantum' photon state, respectively: $|\alpha, \nu\rangle = |\beta, \nu'\rangle = \left|\sqrt{\nu_0}, 0\right\rangle = \left|\sqrt{\nu_0}\right\rangle$ (Figs. 1b,c); $|\alpha, \nu\rangle = |\beta, \nu'\rangle = |0, \nu_0\rangle = |\nu_0\rangle$ (Figs. 1d,e). Also, these examples correspond to the electron energy transfers (i.e., Eq. 2) as we discussed in the previous two sections.

Now we describe the electron coupling process with classical-like photon system in the scheme of weak measurement [10,17], which is given by

$$\underbrace{\langle \beta, \nu' |}_{\text{post-selection}} \underbrace{\mathcal{T} \exp\left( -\frac{i}{\hbar} \int_0^{L/v_0} H_I(t)\, dt \right)}_{\text{electron-photon coupling}} \underbrace{|\alpha, \nu\rangle \otimes |\psi\rangle}_{\text{pre-selection}}$$

$$= \langle \beta, \nu' | \left( 1 + \frac{ie}{\gamma_0 m \hbar} \int_0^{L/v_0} \boldsymbol{A}(t) \cdot \boldsymbol{p}\, dt \right) |\alpha, \nu\rangle \otimes |\psi\rangle$$

$$= \langle \beta, \nu' | \alpha, \nu \rangle \otimes \left| \psi \left( z + \frac{e}{\gamma_0 m} \int_0^{L/v_0} A_w(t)\, dt \right) \right\rangle$$

$$= e^{-|\beta-\alpha|^2/2} |\psi(z - \Delta z)\rangle$$

where we employed the relation $\langle \beta, 0 | \alpha, 0 \rangle = e^{-|\beta-\alpha|^2/2}$ for coherent states with real numbers $\alpha, \beta$. The measuring electron pointer is assumed to be a Gaussian wavepacket in coordinate space ($z$) corresponding to its momentum component (Eq. 1, i.e., $|\psi\rangle$). The final spatial shift of the electron pointer is thus $\Delta z = -\frac{e}{\gamma_0 m} \int_0^{L/v_0} A_w(t)\, dt$ and the corresponding momentum transfer is approximated instantaneously as $\Delta p = \gamma_0 m \left( \frac{\Delta z}{\Delta t} \right) = -e\bar{A}_w$, which confirms the equivalence between the quantum wavepacket momentum transfer in the point-particle limit and the time-averaged weak value of vector potential in the short-time approximation. *[Footnote 2: Note that the weak value $A_w$ is in general a complex number. Conceptually, weak values appear to suggest an approach of describing quantum systems with two boundary conditions (pre- and post-selection), see e.g. [17, 18]. However, $A_w$ is real in our case because the photon state is a coherent state, an eigenstate of the vector potential, thus we expect that $A_w = 2Re\{A^{(-)}\}$, where $A^{(\mp)}$ are negative and positive frequency components of the vector potential, respectively.]*

**Experimental verification.** To verify the two measurement limits from quantum to classical, let us compare our theory with specific experimental results of PINEM in a UTEM and electron spectra in DLA. The two setups depicted in Figs. 4a and 4b (see also [19-22]). Let us clarify that from the point of view of our theoretical one-dimensional model there is no difference between the prism-based setup of (see the inset of Fig. 4c) and the periodic rod structures of Fig. 4b. The relevant classical light field $E(z,t)$ is in the first case the evanescent near field of the laser-illuminated dielectric prism, and, in the second case, a slow-wave space harmonic of a Floquet mode in the periodic rods of the DLA structure. The derived extended electron-photon interaction is the same in both setups as long as the electron and the wave are synchronized or slightly desynchronized with the same detuning parameter $\bar{\theta}$ [9]. Here we mention recent work [13, 23] that, from a different view of photons, studied the quantum-to-classical transition of the photon statistics of quantum light sources on free-electron-light interactions in a UTEM setup with the DLA device.

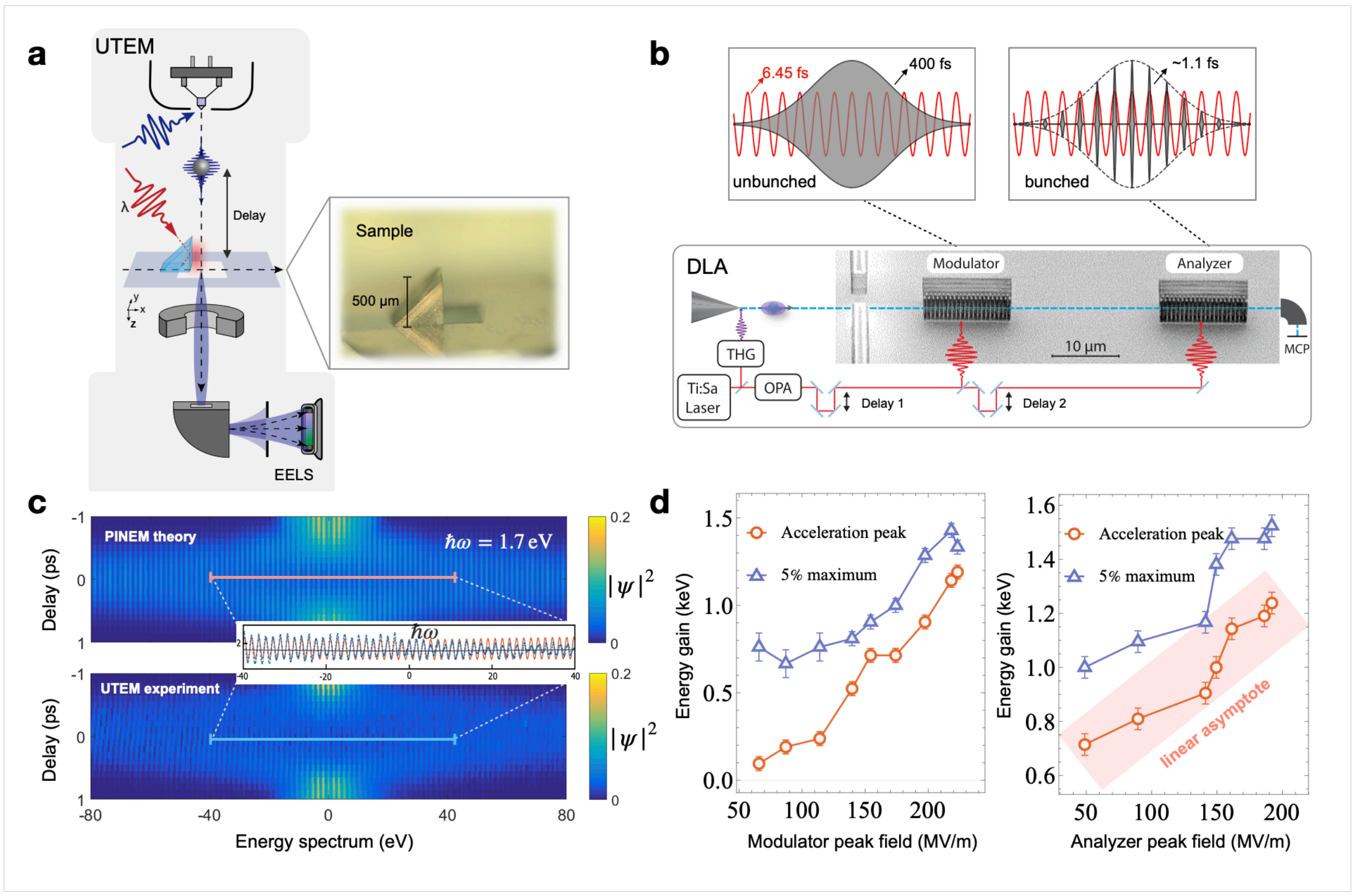

**Fig. 4: Experimental demonstrations of free-electron–photon measurements in the quantum (PINEM) and the classical (DLA) cases.** (a) The schematic setup of UTEM with a prism-mediated laser-near-field electron interaction used to obtain the PINEM spectrum. (b) The schematic setup of the DLA-based experiment with two gratings (modulator and analyzer) used to obtain net electron acceleration. (c) The symmetric energy spectrum with sideband spacing of $\hbar\omega$ (1.7 eV) indicating the quantum regime of the electron-photon interaction. The data (bottom panel) matches well with the PINEM theory of wide electron in plane-wave limit, simulation results of which are shown in the top panel. (d) The left panel shows the net acceleration as function

of the modulator peak field and for a constant analyzer peak field of 207 MV/m. Fig. S2 of the supplementary information shows simulation data with a similar behavior, hence this behavior shows the transition from unbunched to bunched electron beams. With a maximum modulator field strength of 224 MV/m here, the bunched pulse duration is around 1.1 fs, considerably smaller than the optical cycle (6.45 fs), and thus showing the point-particle classicality of the bunched electrons. The right panel shows the energy gain for both the acceleration peak and the maximum 5% (see SI) for a modulator peak field of 224 MV/m. Both depend linearly on the analyzer's field strength demonstrating the weak value measurement of the photon system. The "5% maximum" is the maximum peak acceleration at 5% of the count rate.

In the first case (Fig. 4a), the measured spectrum of the interacting wide electron wavepacket depicts a pattern of multiple discrete sidebands with $\hbar\omega$ spacing. This is explained in terms of PINEM theory in the multiphoton coupling regime (Sec. E in the SI file). Indeed, the width of the measured spectrum with the discrete photon sidebands can be extended to the range of 1keV, and shows no net linear acceleration, even in the case of extremely strong field and non-perturbative laser-electron interaction.

We compare the interaction with the prism in Fig. 4a to the interaction with the second periodic rod structure (the analyzer) in Fig. 4b. The crucial difference is that in the latter case the electrons arrive at the interaction region in a density-modulated state: The optical near-fields in the first periodic rod structure (the modulator) led to an energy modulation without net energy transfer. This together with the free drift of the electron wavepackets generates a train of attosecond-bunched electrons when they arrive at the analyzer (see also [20]). The periodically bunched electrons display a minimal bunch duration of ~1.1 fs, which is smaller than the synchronized optical cycle (6.45 fs). Hence, this situation corresponds to an optical system interacting with a train of electron measuring pointers. In this case, the analyzer can perform a weak measurement of the optical field system when the modulator pre-bunches the electrons, and thus, classical acceleration is expected for the bunched electrons, as shown in Fig. 4d. We can interpret this as a weak measurement of the optical field system as proposed above. The right panel of Fig. 4d shows the linear dependence of the energy gain as a function of the analyzer's field strength for the optimally bunched electrons. We can interpret this as the modulator creating a transition from projective to weak measurement. Furthermore, like in the DLA setup, we can suggest an extended PINEM experiment to have two-stage near-field interactions phase-matched such that the first stage produces attosecond bunching at the second stage, where then large acceleration may be also observed. Additionally, spectrograms of the modulated electron density as a function of delay time are given in Fig. S3.

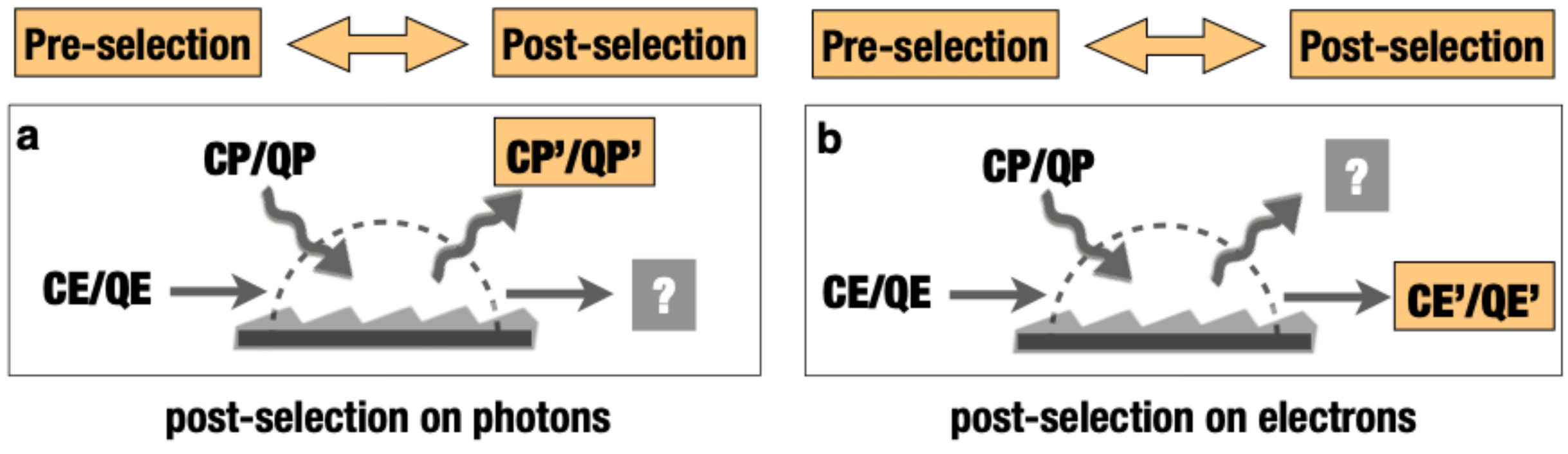

**Fig. 5: The weak-valued electron-photon interaction with pre-/post-selection on photons and electrons.** The pre- and post-selections are performed on (a) the photons or (b) the electrons as the measured system, and the rest acts as the measuring pointer in quantum-to-classical measurement schemes.

**Post-selection on electrons or photons.** Let us discuss the post-selection of the electron-photon states after interaction in terms of measurement theory. Two types of 'weak-valued' electron-photon couplings are schematically shown in Fig. 5. In the reciprocal system-pointer setup of light-matter interaction, the electron can be the measured system, and the photon is then the measuring pointer. If we are able to pre- and post-select the electron wavefunction, detection of the photon radiation rate ($\Delta\nu$) then leads to a shift of the photon pointer, being the measuring pointer, as compared to the measurement of the momentum operator of the electron. In a recent work [9, 24], the reciprocal relation between photon radiation and electron acceleration is demonstrated to be $\Delta\nu + \Delta E/\hbar\omega = 0$, which brings a correspondence between electron spectrum and photon spectrum that conserves the photon exchange in all measurement schemes. This 'Acceleration/Radiation Correspondence' (ARC) relation [9], connects the final measurements of the photon and electron spectrum with/without post-selection as a demonstration of the 'system-pointer' dualism. This setup of weak measurements resembles the pre- and post-selection of atomic states coupled with photons as proposed recently by Aharonov *et al.* [25].

Note that our quantum-to-classical measurement theory is entirely different from the environment-induced decoherence program [26-28]. Decoherence theory, in which the 'classicality' emerges from the natural loss of quantum interference by 'leakage' into the environment [27], does not comprise the contributions of quantum interference, and would neither yield wavepacket-dependent acceleration nor periodic density bunching in the attosecond scale as in [29-31]. Likewise, the environment-induced decoherence cannot produce the classical linear particle acceleration *[Footnote 3: Unlike projective measurements, weak measurements maintain the coherence of quantum states [32,33]. Therefore, one would naively expect that the results agreeing with classical intuition would be obtained in a strong projective measurement, after averaging over all outcomes. For instance, average position of a quantum harmonic oscillator in a coherent state reproduces the evolution of a classical harmonic oscillator. In our analysis, we have demonstrated quantum-to-classical transitions in weak measurements too.]*

**Conclusion**. Four kinds of measurement setups of electron-photon interactions were considered in detail, loosely corresponding to 'classical electron - classical photon', 'classical electron - quantum photon', 'quantum electron - classical photon', and 'quantum electron - quantum photon'. We captured all these interaction types using our unified framework of measurement transition theory, defining all the physics above as a consequence of weak measurement or projective measurement. Then, the transition process was characterized by a universal factor $e^{-\Gamma^2/2}$, which could quantitatively verify our measurement theory in any experiment exhibiting light-matter interactions. Furthermore, our work experimentally reveals the continuous transition from weak to projective measurements, which can also explain the quantum-to-classical transition in common schemes like DLA and PINEM [14].

In addition, we identified the classical linear point-particle acceleration as the weak value of the vector potential and connected it with the appearance of 'classicality' in quantum mechanics. This indicates that weak measurements not only reveal 'anomalous' quantum features of quantum physics, but also surprisingly describes classical characteristics in the realm of classical electrodynamics. The weak value of the vector potential under suitable pre- and post-selections offers a compelling theoretical framework for investigating the interaction between electron wavefunctions and quantum light sources such as superposition of Fock states or squeezed states of light [13, 23]. In a recent paper [34] we have considered, both theoretically and experimentally, the weak-to-strong transition of quantum measurements in trapped ions as a consistent extension of our theoretical framework to generic system-pointer interactions.

## Methods

**UTEM setup** - The measurement of PINEM spectra is conducted on a laser-excited right-angle prism sitting in an ultrafast transmission electron microscope (UTEM). The experimental setup is detailed described in the previous report [18]. Briefly, the UTEM (JEOL-2100P) is operated in nano-beam diffraction (NBD) mode with 70 um condenser aperture and 207.2 kV electron acceleration voltage for small and parallel electron probe. The right-angle prism (BK7 glass, n = 1.512 at 730 nm wavelength) is placed on TEM holder inside UTEM, with one of its right-angle faces aligned parallel to the electron beam. To generate the light evanescent field on the prism face, a 730 nm laser beam from a pump femtosecond laser system (Light Conversion) is coupled to the UTEM from a transparent port through an in-column mirror, and finally incident on the prism. Such beam yields phase-matching of electron and evanescent field with a Cherenkov angle of 19.8 deg. In one acquisition, photon-excited electron wave-packets are produced from cathode illuminated by ultraviolet femtosecond laser pulses from frequency conversion of the pump laser and interacts with the evanescent field. Their energy spectra are captured by an electron energy loss spectrometer (EELS) with many minutes integration time. By changing the delayed time of electron and laser pulses, the two-dimensional map showing the evolution of the interaction as a function of time are revealed as in Fig. 4c.

**DLA setup** – The ballistic bunching is measured in an ultrafast scanning electron microscope. The pulsed electron beam, triggered from a standard Schottky-emitter by a 167 kHz frequency quadrupled Ytterbium fiber laser, and focused to sub-100 nm to enter the channel of the first dielectric laser accelerator structure, the modulator. The DLA structures are fabricated from silicon single crystals via deep reactive ion etching (DRIE). The structures tailor the near field when illuminated by the same fiber laser, upconverted to ~2μm wavelength via an optical parametric amplifier system, to be synchronous with the electrons. The near field interacts in the modulator with the flying electrons and modulates their energy sinusoidally. The subsequent drift between the two structures causes ballistic bunching, due to the non-relativistic electrons having different velocities. The spatial distribution is probed in the second structure, the analyzer. The near field of the analyzer interacts with the modulated attosecond bunched electrons, which leads to a net acceleration. Finally, the energy transfer is revealed with a magnetic spectrometer. More details can be found in [19, 20].

The theoretical modeling, calculations, and supplementary figures and data are available in the SM file.

**Acknowlegements**

We thank Eilon Poem for helpful discussions, and we also thank Zhaopin Chen and Qingqing Cheng for their help to improve the figures. The work was supported in part by the Israel Innovation Authority Grants No. 70002 and 73795, FQXi Grant No. 224321, Pazy Foundation, Israeli Ministry of Science and Technology, Canada Research Chair and Ontario's Early Researcher Award, DIP (German-Israeli Project Cooperation), ISF (Israel Science Foundation) Grant No. 00010001000, I-CORE–Israel Center of Research Excellence program of the ISF, by the Crown Photonics Center, Gordon and Betty Moore Foundation grant 4744 (ACHIP), and ERC Advanced Grant 884217 (AccelOnChip).

**Competing financial interests**

The authors declare no competing financial interests.

Supplementary Materials for

# Demonstration of projective measurements, weak measurements, and quantum-to-classical transitions in ultra-fast free electron-photon interactions

Yiming Pan[1,6†], Eliahu Cohen[2†], Ebrahim Karimi[3], Avraham Gover[4], Norbert Schönenberger[5], Tomáš Chlouba[5], Kangpeng Wang[6], Saar Nehemia[6], Peter Hommelhoff[5], Ido Kaminer[6] and Yakir Aharonov[7,8]

1. Department of Physics of Complex Systems, Weizmann Institute of Science, Rehovot 7610001, Israel
2. Faculty of Engineering and the Institute of Nanotechnology and Advanced Materials, Bar Ilan University, Ramat Gan 5290002, Israel
3. Department of Physics, University of Ottawa, Ottawa, Ontario, K1N 6N5, Canada
4. Department of Electrical Engineering Physical Electronics, Tel Aviv University, Ramat Aviv 6997801, Israel
5. Department of Physics, Friedrich-Alexander Universität Erlangen-Nürnberg (FAU), Staudtstraße 1, 91058 Erlangen, Germany
6. Department of Electrical Engineering, Technion – Israel Institute of Technology, Haifa 3200003, Israel
7. School of Physics and Astronomy, Tel Aviv University, Ramat Aviv 6997801, Israel
8. Institute for Quantum Studies, Chapman University, Orange, CA 92866, USA

†Correspondence and requests for materials should be addressed to E.C. (eliahu.cohen@biu.ac.il) and Y.P. (yiming.pan@campus.technion.ac.il).

**This PDF file includes:**

Supplementary Text
Figs. S1 to S3

## Supplementary Text

### A. The modified 'relativistic' Schrödinger equation from the Klein-Gordon equation

Some of the electron-radiation interaction schemes referred to in the paper (Smith-Purcell radiation, PINEM, FEL, etc.) operate with a relativistic beam; therefore, the use of Schrödinger equation would not be satisfactory for all cases of interest. Since spin effects are not relevant for the present problem, we do not need to use Dirac equation, but rather base our analysis on the Klein-Gordon (KG) equation. Furthermore, following Ref. [11, 9] in the text, we reiterate the derivation of a Schrödinger-like equation out of the KG equation, using the second-order iterative expansion of the free electron energy around its center energy $\varepsilon_0 = \sqrt{p_0^2 c^2 + m^2 c^4}$ . This expansion reduces the quadratic KG equation into the parabolic Schrödinger equation under the well-satisfied approximation that the initial momentum spread and the momentum change due to the interaction are within the range $\Delta \mathrm{p} \ll \mathrm{p}_0$ .

The Klein-Gordon equation originates from the relativistic energy-momentum dispersion:

$$E_p^2 = p^2 c^2 + m^2 c^4, \tag{A1}$$

where $m$ is the electron rest mass, and $c$ is the speed of light. To obtain the KG equation, we make the replacements $E \rightarrow i\hbar \frac{\partial}{\partial t}, \mathbf{p} \rightarrow -i\hbar\nabla - e\mathbf{A}$ （minimal coupling with electromagnetic radiation） and apply the differential operator on a wavefunction:

$$\left(i\hbar \frac{\partial}{\partial t}\right)^2 \psi(r,t) = c^2 \left(-i\hbar\nabla - e\mathbf{A}\right)^2 \psi(r,t) + m^2 c^4 \psi(r,t), \tag{A2}$$

where $e$ is an electron charge. The KG equation can describe the relativistic electrons in most of the considered radiation schemes if spin effects are negligible. If the radiation field is weak, $e\mathbf{A}/mc \ll 1$, then excitation of the negative (positron) energy brunch of the dispersion equation is negligible and one can approximate the wavefunction $\psi(r,t)$ with a single quasi-harmonic positive energy wave

$$\psi(r,t) = u(r,t) e^{-i\varepsilon_{p0} t/\hbar}, \tag{A3}$$

where $\varepsilon_0=\sqrt{p_0^2c^2+m^2c^4}=\gamma_0 mc^2$, $p_0$ the center momentum and $u(r,t)$ is a slowly varying function of time. Then substitution of Eq. (A3) in (A2) and canceling the fast-varying coefficient $e^{-i\varepsilon_0 t/\hbar}$, result in

$$i\hbar\frac{\partial u(r,t)}{\partial t}=\left(\frac{c^2\left(-i\hbar\nabla-e\mathbf{A}\right)^2+\left(m^2c^4-\varepsilon_0^2\right)}{2\varepsilon_0}\right)u(r,t)+\frac{\hbar^2}{2\varepsilon_0}\frac{\partial^2 u(r,t)}{\partial t^2}. \qquad \text{(A4)}$$

This is an exact expression for the slow part function $u(r,t)$. The first-order approximation of its time derivative is

$$i\hbar\frac{\partial u(r,t)}{\partial t}=\left(\frac{c^2\left(-i\hbar\nabla-e\mathbf{A}\right)^2+\left(m^2c^4-\varepsilon_0^2\right)}{2\varepsilon_0}\right)u(r,t).$$

Iterative substitution of this equation into the exact formula (A4) results in

$$i\hbar\frac{\partial u(r,t)}{\partial t}=\left(\frac{c^2\left(-i\hbar\nabla-e\mathbf{A}\right)^2+\left(m^2c^4-\varepsilon_0^2\right)}{2\varepsilon_0}\right)u(r,t)-\frac{1}{2\varepsilon_0}\left(\frac{c^2\left(-i\hbar\nabla-e\mathbf{A}\right)^2+\left(m^2c^4-\varepsilon_0^2\right)}{2\varepsilon_0}\right)^2u(r,t). \qquad \text{(A5)}$$

Now the Klein-Gordon equation can be re-expressed in the form of a modified Schrödinger equation,

$$i\hbar\frac{\partial \psi(r,t)}{\partial t}=i\hbar\frac{\partial u(r,t)}{\partial t}e^{-i\varepsilon_0 t/\hbar}+\varepsilon_0 u(r,t)e^{-i\varepsilon_0 t/\hbar}=H\psi(r,t), \qquad \text{(A6)}$$

where the effective Hamiltonian is

$$H=\varepsilon_0+\left(\frac{c^2\left(-i\hbar\nabla-e\mathbf{A}\right)^2+\left(m^2c^4-\varepsilon_0^2\right)}{2\varepsilon_0}\right)-\frac{1}{2\varepsilon_0}\left(\frac{c^2\left(-i\hbar\nabla-e\mathbf{A}\right)^2+\left(m^2c^4-\varepsilon_0^2\right)}{2\varepsilon_0}\right)^2. \qquad \text{(A7)}$$

The Hamiltonian can be split into an unperturbed electronic part and a radiative perturbation part $H=H_0+H_I(t)$, where

$$H_0\simeq\varepsilon_0+\mathrm{v}_0\left(-i\hbar\nabla-\mathrm{p}_0\right)+\frac{1}{2\gamma_0^3 m}\left(-i\hbar\nabla-\mathrm{p}_0\right)^2, \qquad \text{(A8)}$$

and to first order in the vector potential $\mathbf{A}$,

$$H_I(t) = -\frac{e\left(\mathbf{A}\cdot(-i\hbar\nabla)+(-i\hbar\nabla)\cdot\mathbf{A}\right)}{2\gamma_0 m}, \qquad \text{(A9)}$$

where $\mathrm{v}_0 = p_0/\gamma_0 m = \beta_0 c$ and $\gamma_0 = 1\big/\sqrt{1-\beta^2}$ is the Lorentz factor, $\beta = \mathrm{v}_0/c$ and $m$ is the free electron mass. Note that the coupling term ($\hat{\mathbf{A}}\cdot\mathbf{p}$) has been widely applied to describe numerous kinds of light-matter interactions, like the Smith-Purcell effect, Cherenkov radiation, and transition radiation, regardless of the grating, tip, foil, and nanostructures. For specification in our one-dimensional electron-photon interaction model in a slow-wave structure (such as a grating), we consider a monochromatic laser field with frequency $\omega$, $\hat{\mathbf{A}} = -\frac{1}{2\mathrm{i}\omega}\left(\hat{\mathbf{E}}(\mathbf{r})\mathrm{e}^{-\mathrm{i}\omega t} - \hat{\mathbf{E}}^{\dagger}(\mathbf{r})\mathrm{e}^{\mathrm{i}\omega t}\right)$, where $\hat{\mathbf{A}},\hat{\mathbf{E}}$ are electromagnetic field operators. In our one-dimensional analysis, we assume that the light-electron coupling takes place through an axial slow-wave field component of one of the traveling modes (q): $\hat{\mathbf{E}}(\mathbf{r}) = \sum_{\mathrm{q}} \tilde{\mathrm{E}}_{\mathrm{q}} \mathrm{e}^{\mathrm{i}q_z z - \mathrm{i}\phi_0} \boldsymbol{\varepsilon}_z \hat{\mathrm{a}}_{\mathrm{q}}$, where $\hat{a}_q \left(\hat{a}_q^{\dagger}\right)$ is the annihilation (creation) operator of the photon's Fock state $\left|\nu\right\rangle$ in this quantized mode with wave number $q$ and the normalized polarization vector $\boldsymbol{\varepsilon}_z$ pointing in the direction of propagation $z$.

For the case of our concern,

$$\begin{aligned}
\mathbf{A} &= -\frac{1}{2i\omega}\left(\tilde{\mathbf{E}}(z)e^{-i(\omega t+\phi_0)} - \tilde{\mathbf{E}}^{*}(z)e^{i(\omega t+\phi_0)}\right),\\
\mathbf{E} &= -\frac{\partial \mathbf{A}}{\partial t} = \mathrm{Re}[\tilde{\mathbf{E}}(z)e^{-i(\omega t+\phi_0)}] = E_c\cos\left(q_z z - \omega t - \phi_0\right).
\end{aligned} \qquad \text{(A10)}$$

In our present one-dimensional analysis, we assume a longitudinal field component of a slow-wave structure (e.g. a grating) $\tilde{\mathbf{E}}(z) = E_c e^{iq_z z}\hat{\mathbf{e}}_{\mathbf{z}}$, neglecting transverse field components, and transverse variation of the field. This modified relativistic Schrödinger equation with the effective Hamiltonians (A8), (A9) is used in the main text for the perturbative solution.

## B. The details of the first-order perturbation analysis

Following the standard QED treatment (the time-dependent evolution operator $U = \hat{T}\exp(-i\int H_{int}\,dt')$), we expand the initial wavefunction in terms of the quantum continuous numbers $p$ of the electron state and the Fock number-occupation state of the photon, which is given by $|i\rangle = \sum_{p,\nu} c_{p,\nu}^{(0)} e^{-iE_p t/\hbar} |p,\nu\rangle$, where $c_{p,\nu}^{(0)}$ is the component of the combined electron and photon state $|p,\nu\rangle$ as given in Eq. (1) and $E_p = c\sqrt{m^2c^2 + p^2}$. First-order time-dependent perturbation analysis of the Schrödinger equation results in $i\hbar\dot{c}_{p',\nu'}^{(1)} = \sum_{p,\nu} c_{p,\nu}^{(0)} \langle p',\nu'|H_I(t)|p,\nu\rangle e^{-i(E_p - E_{p'})t/\hbar}$ and the interaction Hamiltonian is taken to be $\hat{H}_I(t) = -\mathbf{e}\hat{\mathbf{A}}\cdot p/\gamma_0 m$ (see Sec. A). By integrating in the time domain to infinity, the emission and absorption processes of the first order perturbed coefficients are given by [9] as follows, $c_{p',\nu'}^{(1)} = c_{p',\nu'}^{(1)(e)} + c_{p',\nu'}^{(1)(a)}, c_{p',\nu'}^{(1)(e,a)} = \frac{\pi}{2i\hbar}\sum_{p,\nu} c_{p,\nu}^{(0)} \langle p',\nu'|H_I^{(e,a)}|p,\nu\rangle \delta\left(E_p - E_{p'} \mp \hbar\omega/2\hbar\right)$, where the matrix elements $H_I^{(e,a)}$ correspond to the emission (e) and absorption (a) parts of the interaction Hamiltonian, respectively.

For energy/momentum transfer in electron energy loss spectrum (EELS), the wavepacket acceleration as the pointer shift is thus obtained as $\Delta E = \sum_{p,\nu} \left|c_{p,\nu}^{(0)} + c_{p,\nu}^{(1)(e)} + c_{p,\nu}^{(1)(a)}\right|^2 \left(E_p - E_0\right)$, where the initial electron energy $E_0 = \sum_{p,\nu} \left|c_{p,\nu}^{(0)}\right|^2 E_p$. Note that the unnormalized final state $c_{p,\nu}^{(f)} = c_{p,\nu}^{(0)} + c_{p,\nu}^{(1)(e)} + c_{p,\nu}^{(1)(a)}$ has the photon-emitted (*e*) and photon-absorbed (*a*) contribution from electron-photon scattering processes $|p,\nu \mp 1\rangle \Rightarrow |p \mp \frac{\hbar\omega}{v_0},\nu\rangle$, respectively. Here we expand together the expressions comprising $\Delta E$, then cancel the initial terms and rewrite as two separate terms: $\Delta E^{(1)} = 2\sum_{p,\nu} \Re\left\{c_{p,\nu}^{(0)*}\cdot c_{p,\nu}^{(1)(e)} + c_{p,\nu}^{(0)*}\cdot c_{p,\nu}^{(1)(a)}\right\}\left(E_p - E_0\right)$ and $\Delta E^{(2)} = \sum_{p,\nu} \left|c_{p,\nu}^{(1)(e)} + c_{p,\nu}^{(1)(a)}\right|^2 \left(E_p - E_0\right)$, where $\Re$ stands for the real part of the argument. The phase-independent term $\Delta E^{(2)}$ is the term that corresponds to photon emission rate, as derived from the Fermi's Golden Rule (FGR) [10-12], while the phase-dependent term $\Delta E^{(1)}$ that originates from quantum interference between the initial state and scattered state is an additional contribution which is usually omitted in the formulation of FGR but leads to the classical linear acceleration [10].

The evolution operator is given by

The quantum recoil of the electron is found from substituting in $i\hbar\dot{c}^{(1)}_{p',\nu'}=\sum_{p,\nu}c^{(0)}_{p,\nu}\langle p',\nu'|H_I(t)|p,\nu\rangle e^{-i(E_p-E_{p'})t/\hbar}$ the energy dispersion relation, expanded to second order: $E_p=c\sqrt{m^2c^2+p^2}\approx\varepsilon_0+\mathrm{v}_0(p-p_0)+\frac{(p-p_0)^2}{2m^*}$, determined by the delta functions and $p_0=\gamma_0 m\mathrm{v}_0, m^*=\gamma_0^3 m$. Then the first-order perturbation coefficients are explicitly given by

$$c^{(1)(\mathrm{e,a})}_{p',\nu'}=\begin{cases}+\left(\dfrac{p'+p^{(0)}_{rec}-\hbar q_z/2}{p_0}\right)\tilde{\Upsilon}\sqrt{\nu'}c^{(0)}_{p',\nu'-1}\mathrm{sinc}\left(\bar{\theta}/2\right)e^{i\left(\bar{\theta}/2+\phi_0\right)}\\-\left(\dfrac{p'-p^{(0)}_{rec}+\hbar q_z/2}{p_0}\right)\tilde{\Upsilon}\sqrt{\nu'+1}c^{(0)}_{p',\nu'+1}\mathrm{sinc}\left(\bar{\theta}/2\right)e^{-i\left(\bar{\theta}/2+\phi_0\right)},\end{cases}\tag{B1}$$

with $p^{(0)}_{rec}=\hbar\omega/\mathrm{v}_0$ and $\bar{\theta}=\left(\frac{\omega}{\mathrm{v}_0}-\mathrm{q_z}\right)\mathrm{L}$ is the classical interaction 'detuning parameter'. The normalized photon exchange coefficient is $\tilde{\Upsilon}=\frac{e\tilde{E}_qL}{4\hbar\omega}$.

To derive the electron energy transfer $\Delta\mathrm{E}=\sum_{\mathrm{p},\nu}\left|\mathrm{c}^{(0)}_{\mathrm{p},\nu}+\mathrm{c}^{(1)(\mathrm{e})}_{\mathrm{p},\nu}+\mathrm{c}^{(1)(\mathrm{a})}_{\mathrm{p},\nu}\right|^2\left(\mathrm{E_p}-\mathrm{E_0}\right)$, the integration over $p$ in (B1) should be carried out with the Gaussian distribution function of the drifted electron amplitude in momentum space: $\mathrm{c}^{(0)}_{\mathrm{p}}=\left(2\pi\sigma^2_{\mathrm{p_0}}\right)^{-1/4}\exp\left(-\frac{(\mathrm{p}-\mathrm{p_0})^2}{4\tilde{\Delta}^2_{\mathrm{p}}(\mathrm{t_D})}\right)\mathrm{e}^{i(\mathrm{p_0L_D}-\varepsilon_0\mathrm{t_D})/\hbar}$ and $\tilde{\Delta}^2_p(t_D)=\Delta^2_{p_0}\left(1+i\xi t_D\right)^{-1}$. For the phase-independent energy transfer emission term $\Delta\mathrm{E}^{(2)}=\sum_{\mathrm{p},\nu}\left|\mathrm{c}^{(1)(\mathrm{e})}_{\mathrm{p},\nu}+\mathrm{c}^{(1)(\mathrm{a})}_{\mathrm{p},\nu}\right|^2\left(\mathrm{E_p}-\mathrm{E_0}\right)$, this involves the following integration:

$$\sum_p \left\{ \left( \frac{p + p_{rec}^{(0)} - \hbar q_z/2}{p_0} \right)^2 p \left| c_{p+p_{rec}^{(0)}}^{(0)} \right|^2 \right\}$$

$$= \left(2\pi\Delta_{p_0}^2\right)^{-1/2} \int dp \left( \frac{p + p_{rec}^{(0)} - \hbar q_z/2}{p_0} \right)^2 \mathrm{p} \exp\left( -\frac{\left(p + p_{rec}^{(0)} - p_0\right)^2}{2\Delta_{p0}^2} \right) \qquad \text{(B2)}$$

$$\approx p_0,$$

and similarly, for the absorption term:

$$\sum_p \left\{ \left( \frac{p - (p_{rec}^{(0)} - \hbar q_z/2)}{p_0} \right)^2 p \left| c_{p-p_{rec}^{(0)}}^{(0)} \right|^2 \right\} \approx p_0. \qquad \text{(B3)}$$

For the phase-dependent energy transfer emission part ( $\Delta \mathrm{E}^{(1)} = 2\sum_{\mathrm{p},\nu} \Re\left\{ \mathrm{c}_{\mathrm{p},\nu}^{(0)*} \cdot \mathrm{c}_{\mathrm{p},\nu}^{(1)(\mathrm{e})} + \mathrm{c}_{\mathrm{p},\nu}^{(0)*} \cdot \mathrm{c}_{\mathrm{p},\nu}^{(1)(\mathrm{a})} \right\} \left( \mathrm{E}_{\mathrm{p}} - \mathrm{E}_0 \right)$):

$$\sum_p \left\{ \left( \frac{p + (p_{rec}^{(0)} - \hbar q_z/2)}{p_0} \right) \Re\left\{ c_p^{(0)*} c_{p+p_{rec}^{(0)}}^{(0)} \right\} \right\} \left( p - p_0 \right) \approx -\left( \frac{\hbar\omega}{\mathrm{v}_0} \right) e^{-\Gamma^2/2}. \qquad \text{(B4)}$$

Analogously, for the absorption term:

$$\sum_p \left\{ \left( \frac{p - (p_{rec}^{(0)} - \hbar q_z/2)}{p_0} \right) \left( c_p^{(0)*} c_{p-p_{rec}^{(0)}}^{(0)} \right) \right\} \left( p - p_0 \right) \approx \left( \frac{\hbar\omega}{\mathrm{v}_0} \right) e^{-\Gamma^2/2}, \qquad \text{(B5)}$$

where we define the decay parameter

$$\Gamma = \left( \frac{\omega}{\mathrm{v}_0} \right) \Delta_{\mathrm{z}}\left(\mathrm{t}_{\mathrm{D}}\right) = \left( \frac{\hbar\omega}{\mathrm{v}_0} \right) \frac{\sqrt{1 + \xi^2 \mathrm{t}_{\mathrm{D}}^2}}{2\Delta_{\mathrm{p}_0}} = \Gamma_0 \sqrt{1 + \xi^2 \mathrm{t}_{\mathrm{D}}^2}, \qquad \text{(B6)}$$

and $\Gamma_0 = \frac{2\pi}{\beta_0} \left( \frac{\Delta_{\mathrm{z}}}{\lambda} \right)$, $\xi = \frac{2\Delta_{\mathrm{p}_0}^2}{\gamma_0^3 \mathrm{m}\hbar}$. Note that in all cases, we used the approximation $\mathrm{p}_{\mathrm{rec}}^{(0)}, \hbar \mathrm{q}_{\mathrm{z}}, \Delta_{\mathrm{p}_0} \ll \mathrm{p}_0$ in the last steps of calculation. Also, note that the imaginary part may contribute to an additional phase within the cosine in the case of very long drift time $t_D$.

## C. The classical electromagnetic field correspondence of the coherent photon state

Within the treatment of quantum electrodynamics, the simplified interaction Hamiltonian in the Coulomb gauge ($\nabla\cdot\hat{\mathbf{A}}=0$) is given by

$$H_I\left(t\right)=-\left(\frac{e}{\gamma_0 m}\right)\hat{\mathbf{A}}\cdot\mathbf{p}, \tag{C1}$$

where the vector potential in second quantization (box quantization with volume V) on a grating is given by (A10)

$$\hat{\mathbf{A}}=\frac{1}{\sqrt{V}}\sum_{q,m}\left(\frac{\hbar}{2\omega_{q,m}\varepsilon_m}\right)^{\frac{1}{2}}\left(\hat{\varepsilon}_q\cdot\hat{a}_{qm}e^{iq_{z,m}z-i\omega_{qm}t}+\hat{\varepsilon}_q^*\cdot\hat{a}_{qm}^{\dagger}e^{-iq_{z,m}z+i\omega_{qm}t}\right), \tag{C2}$$

leading to the electric field

$$\hat{\mathbf{E}}=-\frac{\partial\hat{\mathbf{A}}}{\partial t}=\frac{i}{\sqrt{V}}\sum_{q,m}\left(\frac{\hbar\omega_{q,m}}{2\varepsilon_m}\right)^{\frac{1}{2}}\left(\hat{\varepsilon}_q\cdot\hat{a}_{qm}e^{iq_{z,m}z-i\omega_{qm}t}-\hat{\varepsilon}_q^*\cdot\hat{a}_{qm}^{\dagger}e^{-iq_{z,m}z+i\omega_{qm}t}\right), \tag{C3}$$

where $\hat{\varepsilon}_q$ is the normalized polarization vector (this axial slow-wave field component would be one of the Floquet space harmonics of the radiation mode in the periodic structure of the Smith-Purcell interaction scheme or the axial component in a dielectric structure in Cerenkov radiation interaction schemes). For the given laser field with a longitudinal component $\mathbf{E}=E_c\cos(\omega t-q_z\,\mathrm{z}+\phi_0)\,\hat{\mathrm{e}}_z$, the corresponding coherent state $\left|\alpha\right\rangle$ for the m[th]-order Floquet harmonics is then obtained as

$$\alpha=\sqrt{\frac{\varepsilon_{eff}V}{2\hbar\omega}}\mathrm{E}_c e^{-i(\phi_0+\pi/2)}, \tag{C4}$$

where we consider one harmonic of monochromatic near-field photonic excitation with synchronization condition, $\omega/q_{zm}=v_0$ and the laser frequency is fixed at ω, and we obtain $\mathrm{E_c}=|\alpha|\tilde{E}_q$. We should note that the quantization of near-field excitation is a non-trivial issue that relates to the specific configuration of near-field harmonics distribution and spectrum on the grating and the laser illumination condition. Here for our simplified model, we use the free-space quantization for the interaction light field and naively absorb all the structural information of the near-field into the effective dielectric constant $\varepsilon_{eff}$.

## D. The Wigner function representation of our measurement theory in phase space and its comparison with decoherence theory

The comparison between our measurement theory and decoherence theory can be also slightly explained in Figs. 2a and 2b. Based on the conventional QED formulation, we can obtain three spectral sidebands (i.e., one initial sideband, single-photon-emitted sideband and single-photon-absorbed sideband) in the final electron wavefunction $\psi_p = \psi_0(p - p_0) + \psi_{-1}(p - p_0 - \hbar\omega/v_0) + \psi_1(p - p_0 + \hbar\omega/v_0)$. If we present the final electron wavefunction in phase space as Wigner function,

$$W(p,z) = \frac{1}{\pi}\int \psi_{p+\frac{q}{2}}\psi^*_{p-\frac{q}{2}}\, e^{iqz/\hbar}\, dq \tag{D1}$$

then we can explicitly obtain $W(p,z) = W_{0,0} + W_{1,1} + W_{-1,-1} + 2\Re\{W_{0,1} + W_{0,-1} + W_{1,-1}\}$ from the first-order perturbative approximation. These interference fringes can be viewed explicitly in Fig. 2c. The interference parts between sidebands in the point-particle limit of electron wavepacket leads to the central momentum shift of the final electron Wigner function, as shown in Fig. 2a. However, when considering decoherence, all interference terms are suppressed $2\Re\{W_{0,1} + W_{0,-1} + W_{1,-1}\} = 0$. In this situation, the final electron momentum distribution symmetrically broadens, with no wavepacket acceleration in the spectrum, as similarly shown in Fig. 2b.

It has to be noted that Fig. 2b is the interaction between classical electron and quantum photon ($|\nu\rangle$). Thus, these interference terms do not exist intrinsically because of the orthogonality relation of the associated photon states with the electron sidebands, that is $\langle\nu|\nu \pm 1\rangle = 0$. The natural suppression of quantum interference in the cases 2b and 2d is the same as the prediction from the decoherence program, in which the photon state acts like the environment state.

## E. The projective-to-weak measurement transition of periodically bunched electron beams: the net linear energy transfer of the PINEM spectrum

Here, we will demonstrate the weak measurement of an electromagnetic field with a periodically bunched electron having PINEM photon sidebands (with discrete spectral spacing $\hbar\omega$). In the context of quantum measurement theory, the measuring apparatus is a train of periodically bunched attosecond electron pulses. Considering the free electron laser interaction as having a short duration (neglecting the chirp during the interaction), we can describe the interaction in the following simplified Hamiltonian,

$$H = H_0 + H_{int}, \tag{E1}$$

where the first term is the free electron Hamiltonian (A8), and the second term is the interaction part (C1). We ignore the dynamics of the optical field during the short interaction time. Based on the Schrödinger equation, we shall describe the interaction of an electron with two sequential near-fields in the following mathematical form,

$$|\psi_f\rangle = A(g_2)U(t_D)A(g_1)|\psi_i\rangle, \tag{E2}$$

where $\psi_{i,f}$ are the initial unperturbed and finial modulated free electron wave functions. The evolution operator is divided into three parts with $A(g_1)$ being the interaction on the first modulator, $U(t_D)$ being the free drift between two near-field interactions, and $A(g_2)$ being the interaction on the second analyzer. Without loss of generality, the three operators can be modeled as

$$\begin{aligned} A(g_{1,2}) &= \exp\left(-2i|g_{1,2}|\sin\left(\frac{\omega z}{v_0} + \phi_{1,2}\right)\right), \\ U(t_D) &= \exp\left(-iH_0\left(\frac{\partial}{\partial z}\right)t_D\right). \end{aligned} \tag{E3}$$

with $g_{1,2} = \frac{e}{2\hbar\omega}\int F(z)e^{-i\omega z/v_0}\,dz = 2\sqrt{v_0}\,\widetilde{\Upsilon}$ as we derived previously. The longitudinal near-field $F(z)$ can be described by an effective traveling sinusoidal function [Hommelhoff-PRL-2019, Ropers-NPhoton-2017] (omitting the transverse coordinate). The coupling constants $g_{1,2}$ are complex, and the relative phase difference $|\phi_1 - \phi_2|$ is the time delay (delay 2 in Fig. 4b) between the two interactions. There is a relevant time delay (delay 1 in Fig. 4b) between the initial wave function and the modulator near-field given by $|\phi_1 - Arg(\psi_i)|$. The free electron Hamiltonian in the nonrelativistic limit is given by Eq. A8, where only the longitudinal derivative is involved $\nabla \to \frac{\partial}{\partial z}$.

In principle. the specific electron-photon interaction process as described by (E2) can be solved analytically. In our case, we are curious about obtaining the final accelerated electron wave function numerically with an asymmetric PINEM spectrum. Let us assume the initial electron wave function to be a chirped Gaussian (Eq. 1), given by

$$\psi_i(z,t) = \frac{e^{\frac{i(p_0 z - \varepsilon_0 t)}{\hbar}}}{(2\pi\Delta_{z_0}^2)^{\frac{1}{4}}\sqrt{1+i\xi t}} \exp\left(\frac{(z-v_0 t)^2}{4\Delta_{z_0}^2(1+i\xi t)}\right), \tag{E4}$$

with the initial wave packet waist $\Delta_{z_0} = \hbar/2\Delta_{p_0}$. For the condition that $\Delta_{z_0} \gg \beta\lambda$ corresponds to the quantum plane-wave limit (see Fig. 3d), the spectrum of modulated intermediate electron state $(A(g_1)\psi_i\rangle)$ will produce symmetric photon spectral sidebands. The free drift process would translate the imprinted energy modulation into a periodic charge density modulation, yielding the attosecond bunched electron pulses in the propagated intermediate state $(U(t_D)A(g_1)\psi_i\rangle$ (see the inset of Fig. 4b). The attobunched electron pulse size is much smaller than the wavelength. The periodicity of the pulse trains is synchronized with the optical cycle so that the whole pulse train can be accelerated in a classical fashion. It is a surprising feature because the net energy transfer in the second interaction of the analyzer $(|\psi_f\rangle)$ emerges from the quantum PINEM interaction by exchanging discrete photon quanta.

Fig. S1 demonstrates the density modulation and corresponding energy modulation of the final free electron wave function. Fig. S1a-d shows the final modulated electron after interacting with the modulator near-field without and with free drift, respectively. With the control of the drift duration $(t_D)$, we can produce the optimal charge density bunching (see Fig. S1c). Note that the drifted PINEM spectrum (S1d) is the same as the undrifted case (S1b) because the momentum is a good quantum number in the duration of free propagation. Fig. S1e-h shows the final modulated electron after passaging the interaction region of the analyzer. Figs. S1e and S1f are calibration scenarios where the two interaction regions are too close, and no intermediate free drift occurs. We can easily observe a broader PINEM spectrum in Fig. S1f. The most striking result is the accelerated PINEM spectrum for the periodically bunched electron wave function, as shown in Fig. S1h. The whole energy shift in PINEM stems from the quantum interference between photon sidebands. From a classical perspective, each attobunch experiences the dynamics of point-particle in the presence of a synchronized electromagnetic field. The asymmetric spectral feature validates the possibility of performing a weak measurement of the analyzer near-field via the bunched and synchronized electron pulse trains instead of a single bunched electron.

To assess the classical emergence of weak measurement in the analyzer interaction regime, Fig. S2 demonstrates the linear net acceleration of the periodically bunched pulses trains. Fig. S2a shows that the optimal drift duration is dependent on the modulator near-field strength ($g_1$). For a fixed drift duration (corresponding to the distance between the modulator and the analyzer), the energy gain is maximal at the optimal bunched case, whereas the under bunched and over bunched electron beams cannot be accelerated efficiently. Fig. S2b shows the striking linear dependence of the energy gain as a function of the analyzer near-field strength ($g_2$), even for the under bunched electron pulse trains. The quantum analysis (Fig. S2) is in nicely consistent with the experimental data (Fig. 4d in the main text). The relevant parameters are given in the figures.

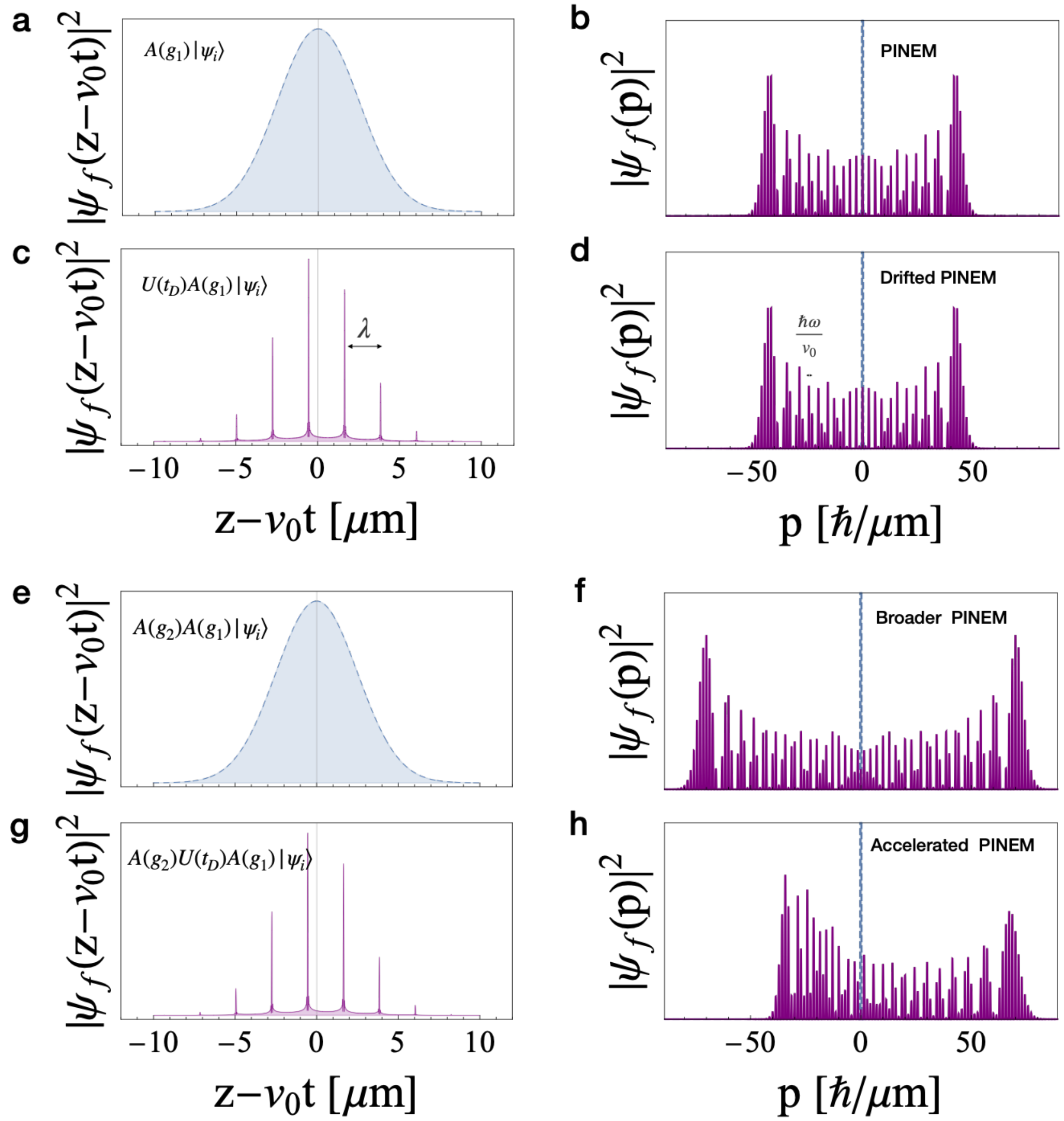

**Fig. S1: The final density modulation and energy modulation of free electron wave function in free-electron and photon interaction.** (a, b) The modulated density and energy distribution after passing the modulator interaction region ($\boldsymbol{A(g_1)\psi_i}\rangle$). (c, d) The modulated density and energy distribution after free drift duration ($\boldsymbol{U(t_D)A(g_1)\psi_i}\rangle$). (e, f) The modulated density and energy distribution after passing the analyzer interaction region ($\boldsymbol{A(g_2)A(g_1)|\psi_i}\rangle$), but without considering the free drift. (g, h) The modulated density and energy distribution after passing the analyzer interaction region ($\boldsymbol{A(g_2)U(t_D)A(g_1)|\psi_i}\rangle$).

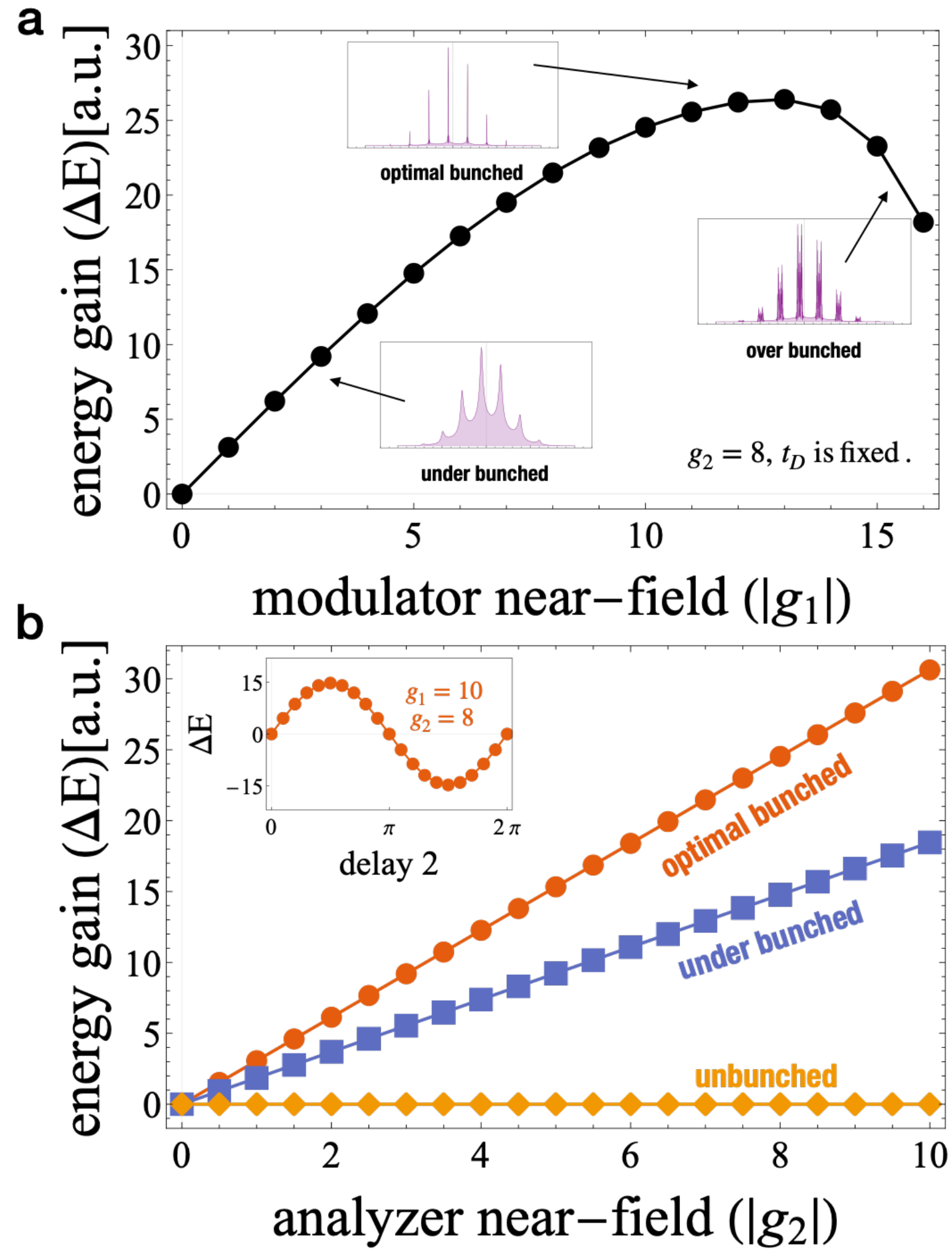

**Fig. S2: The linear net acceleration of periodically bunched electron pulses trains showing the weak measurement of the analyzer near-field.** (a) Numerical simulations of the energy gain of the electron pulses trains having traversed both modulator and analyser stage, as a function of the modulator near-field strength. The inset shows the bunching state at the analyser. The resulting net energy gain is plotted here, for under-bunched, optimally bunched, and over-bunched electron pulse trains. The analyser field strength was kept constant here. (b) The opposite situation: Here, the modulator field strength is kept constant at three different values, indicated by the colors: For an unbunched beam, clearly no net energy gain is expected, while it becomes maximal for the optimally bunched beam (orange). The linear dependence of net acceleration for the bunched electron pulses as a function of the analyzer near-field strength is the main result here, as also seen in the experiment. The inset shows the phase dependence of the point-particle-like acceleration of periodic bunches.

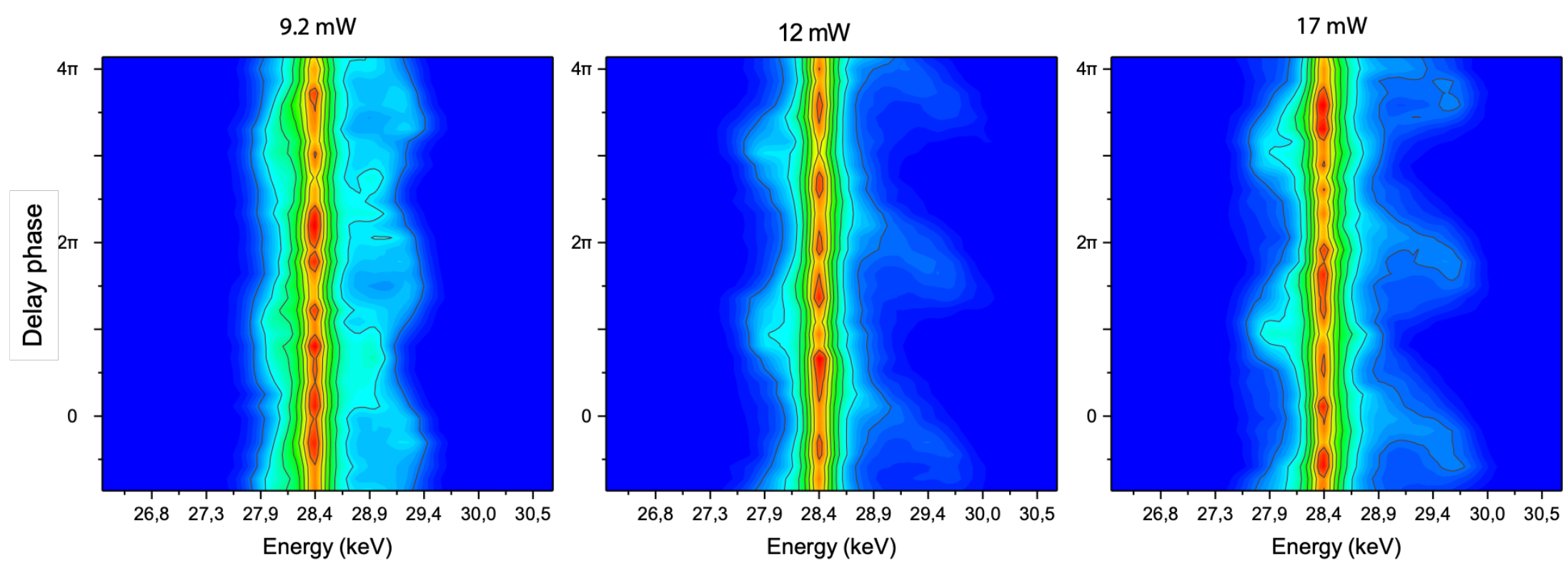

**Fig. S3: Three example spectrograms showing the energy distribution as function of relative phase between laser pulses driving the modulator and the analyser.** The delay axis is presented in units of the optical cycle (6.45fs). Three different modulator powers of 9.2 mW, 12 mW, and 17mW are shown, corresponding to peak field strengths of (141, 161, 191) MV/m. The colour shows electron counts, normalized to the maximum for each data set individually. The "acceleration peak" in Fig. 4 of the main text is found from slices of these data. The "5% maximum" is the maximum acceleration found where 5% of the maximum count rate shows the largest energy gain.